# A Practical Entity Linking System for Tables in Scientific Literature

**Varish Mulwad,**[1] **Tim Finin,**[2] **Vijay S Kumar,**[3]
**Jenny Weisenberg Williams,**[4] **Sharad Dixit,**[5] **Anupam Joshi**[6]

GE Research, Niskayuna, USA [1,3,4,5]
University of Maryland, Baltimore County, Baltimore MD [2,6]
varish.mulwad@ge.com[1], finin@umbc.edu[2], v.kumar1@ge.com[3], weisenje@ge.com[4], sharad.dixit@ge.com[5], joshi@umbc.edu[6]

**Abstract**

Entity linking is an important step towards constructing knowledge graphs that facilitate advanced question answering over scientific documents—including the retrieval of relevant information present in tables within these documents. This paper introduces a general-purpose system for linking entities to items in the Wikidata knowledge base. It describes how we adapt this system for linking domain-specific entities—especially for those entities embedded within tables drawn from COVID-19-related scientific literature. We describe the setup of an efficient offline instance of the system that enables our entity-linking approach to be more feasible in practice. As part of a broader approach to infer the semantic meaning of scientific tables, we leverage the structural and semantic characteristics of the tables to improve overall entity linking performance.

## Introduction

The rapid pace of research in dynamic, fast-evolving scenarios—as recently exemplified by COVID-19 and the unprecedented volumes of scholarly literature on this subject (Else 2020)—has necessitated more machine-driven, human-interpretable approaches to scientific knowledge discovery. Open datasets like CORD-19 (Wang et al. 2020) have motivated novel techniques and tools for keyword/semantic search and Q&A, recommendation, and summarization of scientific documents. As with the web, discovery from scientific literature is predominantly associated with searching over unstructured textual content. Domain-specific neural search engines (Zhang et al. 2020), (Hall 2020) typically produce ranked lists of matching articles in response to search requests, while mainstream information retrieval methods may also deliver direct short, targeted responses (drawn from text) to queries. To facilitate such a search, Sohrab et al. (2020) introduced the BENNERD system and an annotated subset of CORD-19 articles to demonstrate the fundamental tasks of named entity recognition and entity linking for COVID-19-related entities found in text.

Besides text, alternative modalities such as tables and charts have come to play a considerable role in how the scientific community succinctly conveys descriptive information in the literature. Our experience with assembling a corpus of over 62,000 open-access coronavirus-related articles from PubMed Central (PMC 2003) between 2020-21 yielded over 120,000 tables underlining a wealth of latent knowledge embedded within these structured artifacts. The extraction and retrieval of relevant information from these scientific tables is becoming increasingly critical to emerging knowledge-driven applications. For example, consider a genomic surveillance scenario seeking information on *treatment efficacies against the top prevalent COVID-19 variants in each US state*. Better responses to such queries entail going beyond text and searching relevant portions of or entire scientific tables for vital knowledge nuggets, possibly fusing information from multiple source tables on the fly.

Although learning-based representational models for tabular data (Yin et al. 2020) show great promise for understanding relationally structured web tables, these models are typically not tuned to unconventional structural complexity.

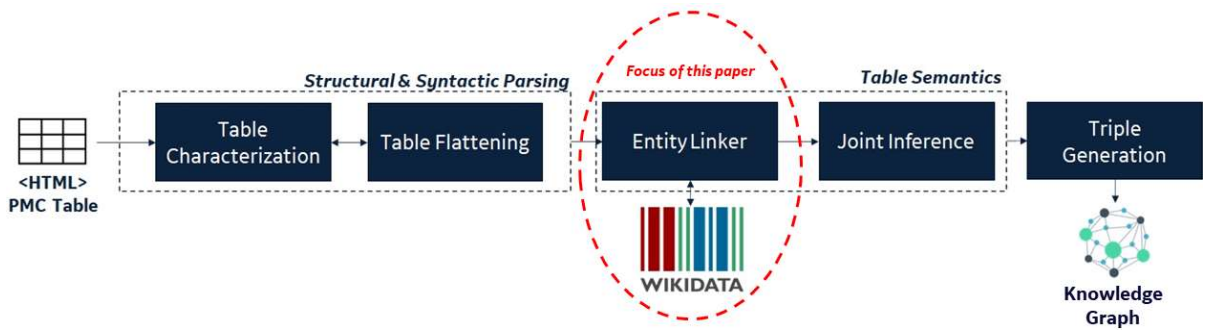

Figure 1. Entity linking and its role in constructing knowledge graphs from scientific tables

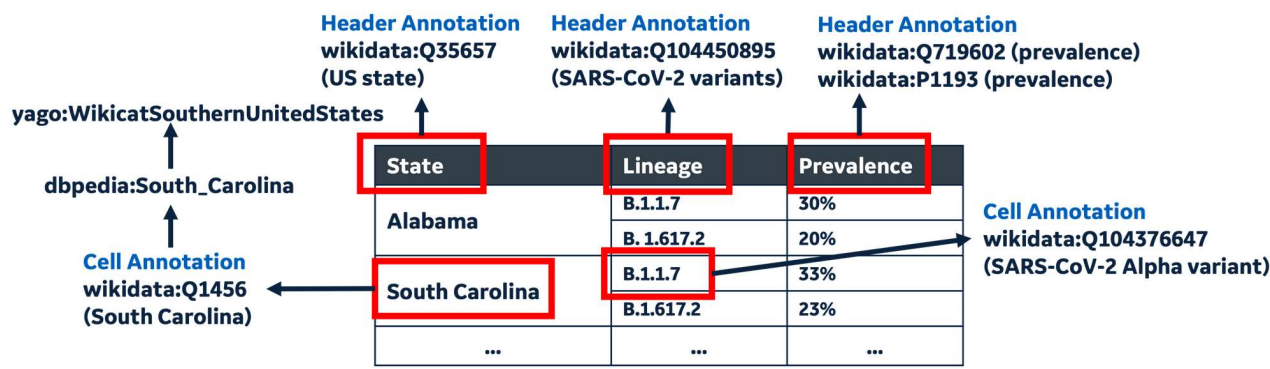

Figure 2. Examples of header and cell annotations of links to Wikidata items and properties

This is especially true for the dense and often implicit semantics and diffuse context inherent in scientific tables in highly specialized domains (Mulwad et al. 2014). Representing scientific tables as semantically annotated linked data artifacts accounts for structural complexities and enables explicit reasoning over tabular content to infer their semantics and relevance to search queries. Hence, entity linking (EL) is fundamental to our end-to-end pipeline for constructing such knowledge graphs of tables drawn from scientific documents, as depicted in Figure 1.

This paper presents an EL system to automatically map the content of individual cells in scientific tables to appropriate entries in the Wikidata knowledge base (Vrandecic and Krotzsch 2014). To keep up with the scientific literature infodemic, we architected a more efficient local, offline EL system using periodic Wikidata knowledge dumps. While the ensuing efficiency gains make our system more feasible in practice, we discuss the implications on EL performance.

## Entity Linking for Scientific Text and Tables

Given a mention of an entity in a document and a unique set of known entities defined in some knowledge base, entity linking refers to finding and assigning the entity ID corresponding to the mentioned entity. Entities play an important role in text and are often used to describe what the text is about. Likewise, linking entity mentions in the header and body cells of tables, as well as linking entities in captions or other referring text can help partly understand or infer the semantic meaning of tables. We developed a general-purpose linker to link entity mentions in text to items in (and to further extract useful information about items from) Wikidata. We describe the linker's customizations and inner workings for linking highly specialized, idiomatic content within header and body cells of tables drawn from a corpus of COVID-19-related scientific literature.

## Wikidata: Reference Knowledge Base

Wikidata (Vrandecic and Krotzsch 2014) is a collaboratively edited multilingual knowledge graph used to provide common data for Wikimedia projects with currently about 1.2 billion facts on over 102 million items. Wikidata's ontology has a fine-grained type system with more than two million types and about 11 thousand properties, including an item's label, aliases, and description. Each Wikidata item has a unique identifier beginning with Q, like Q3519875 ("National Institute of Allergy and Infectious Diseases"), and each property has an identifier starting with P. The property P31 (*instance of*) links an item with its immediate types, P279 (*subclass of*) links a concept item to its immediate super-types, and P1647 (*subproperty of*) links properties to their immediate super-properties.

An entity has just one label in a given language, its "canonical name". An entity can have any number of aliases in a language and can have a short description in any language. Unlike other open knowledge graphs, Wikidata includes and links to specialized knowledge from additional domain-specific knowledge resources. These include the Unified Medical Language System (UMLS) (Bodenreider 2004) knowledge base and the Medical Subject Headings (MeSH) thesaurus (Lipscomb, 2000), which bring together biomedical vocabularies and standards to enable interoperability.

Figure 2 shows an example of a simple scientific table with links to appropriate Wikidata items that highlights several high-level issues we addressed. One is that we must consider the "header" cells (whether for columns or rows) differently from the regular table body cells. Note that the third column's header cell, *Prevalence*, has two good candidate links: the concept Q719602 ("number of disease cases in a given population at a specific time") and the property P1193 ("portion in percent of a population with a given disease or disorder"). We give preference in such cases to using the property item over the concept item.

The middle header cell containing the text *Lineage* illustrates a second issue: A simple linker might choose the most common match for this based only on the text: Q1517820 ("line of ancestors and descendants of a person"). However, the cells in this column (e.g., B.1.1.7) are all easily matched to Wikidata items whose immediate type is Q104450895 ("variant of SARS-CoV-2"). Therefore, we need to do joint inference using both the header cell and a sample of its data cells to choose the best links for both.

The first column of the table highlights a third aspect of the task: mining additional knowledge from resources connected to candidate Wikidata items. Wikidata items often link to other knowledge graphs, such as DBpedia (Bizer et al. 2009), that contain additional useful information. DBpedia, for example, has a short paragraph describing its items and links to types in the Yago fine-grained type system (Suchanek et al. 2007).

**Core Entity Linking Algorithm**

Our entity linker takes a mention string (e.g., from a table header or cell) and begins by retrieving a pre-specified number of Wikidata items using the MediaWiki search API. This returns a ranked list containing each item's Wikidata ID, label, aliases, and English language description. Next, we re-rank candidates to promote ones that resulted in an exact match of the mention string with a Wikidata item's label (best) or alias (second best). For each candidate, we use a SPARQL query to retrieve its types, both immediate (P31) and inherited, via a chain of P279 links for concept super-classes and P1647 links for property super-properties.

For specific domains, our linker leverages the ultra-fine-grained Wikidata type system to infer additional domain types for an item by checking for specific domain-relevant properties. We identified a custom set of Wikidata item types and properties to support EL for the biomedical domain. For example, we infer the *mesh item* type if an item has a MeSH descriptor ID property (P486) that connects the item with a UMLS Medical Subject Heading.

When linking the text in a header cell, we give more weight to candidates that are Wikidata properties. For example, candidates for the text "location" include an item representing the geographic location (Q2221906) as well as the property location (P276). While either might be relevant, our annotation methodology strongly preferred the latter.

The linker's filtering and ranking of candidate items are based initially on an analysis of an item's types. This type analysis is controlled by five lists of types that are part of the linker's configuration for a domain and task. These are ordered from best to worst as follows: **(1)** *Target* types are those we want to find based on the mention type identified by an NLP system; **(2)** *Near-miss* types are close to the target types and often confused with the targets by an NLP system; **(3)** *Good* types are ones that are very relevant to the domain, such as a MeSH term (Medical Subject Heading); **(4)** *OK* types include types that are acceptable and common in many domains, such as organizations, people, geo-political entities, and locations; and **(5)** *Bad* types are ones we are not interested in (e.g., fictional characters, journal articles, musical groups) and result in a candidate being immediately rejected.

The type names of interest are mapped to Wikidata types via the linker's configuration dictionary. Extending this dictionary enabled us to easily customize our linker to specific domains such as COVID-19-related scientific research. For our domain, examples of *good* types are Wikidata high-level classes corresponding to disease, protein, chemical compound, vaccine type, and type of statistic. *OK* types are those associated with the standard OntoNotes (Weischedel et al. 2011) types such as person, event, facility, organization, and location. Entities of these types often occur in biomedical tables. Our *bad* types cover things like songs, works of art, sports organizations, fictional things, and other high-level types unlikely to be present in medical tables. For example, there exist 83 Wikidata items with the canonical name "virus". These include Q808, the infectious agent, as well as films, songs, musical albums, rock groups, painting, video games, musicians, a professional wrestler, and more.

Finally, we have a mapping of *near-miss* types that represent types that are easily confused. A classic example is the OntoNotes types FAC (for facility) and LOC (for location) are easily confused by most NLP systems. An entity like *Wuhan Institute of Virology* can be marked as an ORG, LOC, or FAC depending on its context. Since locations are a common type in tables for this domain, we can treat an item identified as a FAC or ORG by a language processor as possibly referring to a location. Additional ranking for an item's prominence is then done using its number of sitelinks, i.e., the number of links to other Wikimedia projects that contain information about the item.

Beyond type analysis-based filtering, the last step is the ranking of the final candidates using a context span or string, if provided. The similarity of the context and the item's description is computed with embeddings from the spaCy (Honnibal 2020) large language model and generates a score that is used along with the item's rank in the candidate list to select and return the best link. This worked reasonably well for both well-structured text (e.g., table captions) and for collections of terms from the row and column headers and could be improved by using an embedding model fine-tuned on the biomedical domain.

**Efficient Entity Linking at Large Scale**

Our entity linker initially used the Wikidata and Wikimedia APIs to retrieve the initial ranked list of Wikidata candidate items and their type and supertype information. Since Wikidata is a public resource, the APIs are understandably rate-limited such that unreasonable access requests and query rates in excess of established limits may lead to IP address blacklisting (WDQS 2022). Table 1 breaks down our average observed entity linking time to link a single exemplar mention string to a Wikidata entity while operating under the above limits. Accessing public Wikidata APIs, our linker can operate no faster than around 30 seconds per entity. For our dataset of 120,000+ tables (a rate reflective of the COVID-19 infodemic), annotating even just 10 cells per table at this rate could end up taking over a year.

| Entity Linking algorithm stage | Time |
|---|---|
| Retrieve initial ranked list of candidate items via Action API (top 20 candidates) | 12 sec. |
| Retrieve all types and supertypes per candidate via WDQS. Filter and rank candidates based on type analysis (good/OK/bad) for domain | 18 sec. |
| Similarity computation based on context string | N/A |
| **TOTAL** entity linking time per mention string | **30 sec.** |

Table 1. Entity linking time using Wikidata APIs

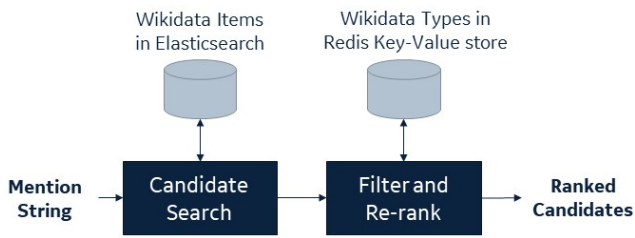

Figure 3. Functional architecture of an efficient 'offline' entity linker

Furthermore, when applying EL to infer table semantics (see next section), linking of a single header cell could in turn translate to linking of all other cells in the respective column or row—potentially placing far greater stress on the linker. As a result, while Wikidata APIs facilitated a proof of concept of our core entity linking algorithm, they cannot sustain a practical, scalable linking service capable of keeping up with contemporary scientific publication rates.

To address these API rate-limit bottlenecks, we initially set up a transient caching layer for cell entity linking results so that future requests to link the same mention string would be served from the cache, avoiding API invocations. However, this strategy was insufficient, so we decoupled our core entity linker from the public Wikidata altogether by architecting and progressively setting up a more efficient system using local periodic dumps of relevant Wikidata knowledge.

The system is offline in that the linker no longer relies on Wikidata APIs. Wikidata's complex software architecture (WD Architecture 2018) and its enormous size make it challenging to replicate locally in its entirety. That said, our entity linker does not need all capabilities that Wikidata offers. We targeted emulation strategies addressing bottlenecks with cross-item graph search (via WDQS and Wikidata's underlying RDF triple store), and full-text search over items and its properties (via the Action API and underlying CirrusSearch Wikibase extension). We leverage proven open-source storage technologies such as the Elasticsearch engine and the Redis key-value store to emulate underlying Wikidata capabilities, as depicted in Figure 3.

We implemented this system by uploading partial JSON dumps of Wikidata items, their basic attributes (label, aliases, description), specific types, and 'sitelinks' counts into a local Elasticsearch index. This resulted in a locally searchable collection of 95.8M items. Offline, we retrieved the current type hierarchy (by traversing P31 and P279 property relationships) and loaded the resulting dictionary mapping each of Wikidata's 2.6M types to its supertypes into Redis. This reduced determining if an entity was an instance of a given type (direct or inherited) to a dictionary lookup.

In this efficient EL system, an initial candidate search is performed using an Elasticsearch multi-match query that compares a mention string against labels and aliases. In lieu of Wikidata's CirrusSearch ranking mechanisms, we use an item's sitelinks count (i.e., popularity) as a proxy for its prominence and rank candidates in descending order of their sitelinks counts. Once we have a ranked list of candidates for each item, we query Redis using the item's entity ID and direct types as keys to retrieve associated inherited types. Type analysis and re-ranking then proceeds as before.

Figure 4 shows a progression in our replacement of Wikidata API invocations with queries to these local knowledge stores. The resulting system trades linking accuracy for a three-fold improvement in linking efficiency, with potential for even further speedups via parallel processing. The impact on entity linking performance is largely dictated by the quality of the initial ranked candidate list returned by our Elasticsearch query. We are exploring techniques such as PageRank to better estimate an item's relative importance.

## Entity Linking to Infer Semantics of Tables

The meaning of text derives from its constituent words, which in turn are understood using grammatical knowledge and context provided by surrounding text. Inferring the intended meaning of tables additionally requires interpreting row/column headers, and relations between them, besides linking cell values to entities. To improve EL performance for inferring semantics of scientific tables, we supplement our core algorithm with other techniques (beyond the scope of this paper) as shown in Fig. 1. These include:

- Rule-based syntactic characterizations that categorize tables into types (e.g., horizontal) based on their structure;
- Joint inference using embeddings of Wikidata items and Wembedder-driven (Nielsen 2017) clustering operations

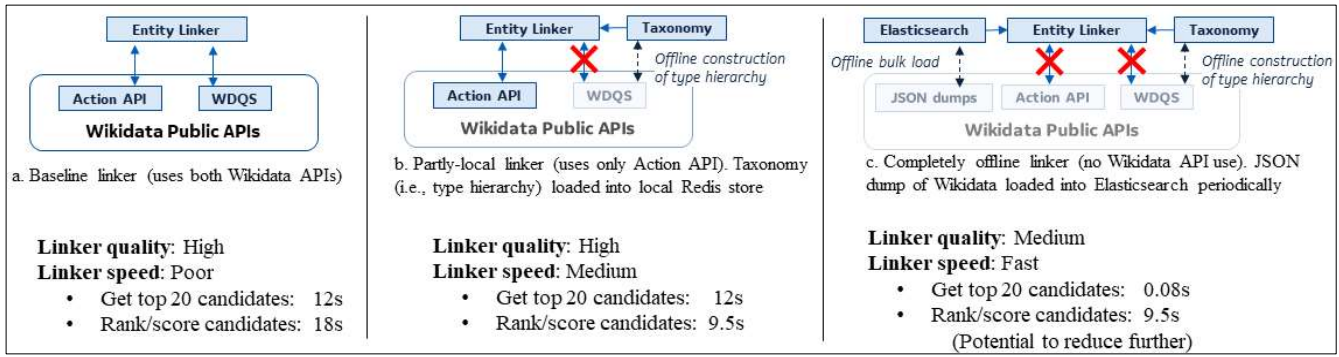

Figure 4. Replacing the entity linker's use of public Wikidata APIs with efficient offline, local queries

to compute compatibility between entities and assign entities to cells in a column; and

- Specialists using pattern-based or machine-learning approaches to assess commonly encoded data types in cells to avoid linking their values for specific kinds of literals, e.g., RNA/DNA sequences or Clinical Trial IDs.

Our EL system achieves a fair degree of accuracy in linking table cells to Wikidata items. We based our evaluations on a manually annotated subset of 47 tables extracted from 45 COVID-19-related articles drawn at random from PubMed Central (PMC 2003). Of the 910 table cells (out of a total of 3600 manually annotated cells in these tables) expected to be mapped to a Wikidata item, our linker achieved a recall of 0.82 when the expected annotation was part of the linker's initial candidate item set, and a precision of 0.51 over the subset of these cells with expected Wikidata annotations.

## Discussion and Conclusions

Existing NLP tools for entity linking (Honnibal 2020) support a very limited entity type system, often based on just OntoNotes types (e.g., PER, ORG, LOC, FAC) and do not cover specialized scientific entities. The SemTab challenge on Tabular Data to Knowledge Graph Matching focused on three mapping tasks aimed at inferring the semantics of web tables (Jimenez-Ruiz et al. 2020). While it recently included tables from biology literature, leading tabular EL systems (Chabot et al. 2019) do not adequately cover domain-specific entities. Bespoke EL systems for COVID-19-related entities (Sohrab et al. 2020) link against UMLS and do not use Wikidata's extensive type hierarchy or entity coverage.

Part of our goal is to fill this missing gap with a practical EL system that can not only be adapted for domain-specific entities but can also help infer table semantics with high accuracy by leveraging Wikidata's rich type system. As entity linking of tables against Wikidata at large scale is bottlenecked by rate-limited APIs (Nguyen and Takeda 2022), we built an offline version of our EL system, achieving a threefold improvement in efficiency while sacrificing a tolerable reduction in linking performance.

## Acknowledgments

This research is based on work supported in part by the Office of the Director of National Intelligence (ODNI), Intelligence Advanced Research Projects Activity (IARPA), via [2021-21022600004]. The views and conclusions contained herein are those of the authors and should not be interpreted as necessarily representing the official policies, either expressed or implied, of ODNI, IARPA, or the U.S. Government.